\documentclass[a4paper]{article}
\usepackage[margin=25mm]{geometry}
\usepackage{amsfonts}
\usepackage{amssymb}
\usepackage{graphicx}
\usepackage{svg}
\usepackage{amsmath}
\usepackage{caption}
\usepackage{etoolbox}
\usepackage{bm}
\usepackage[labelfont={small,bf}]{caption}
\usepackage[font={small}]{caption}
\usepackage{subcaption}
\usepackage{hyperref}
\usepackage{authblk}

\title{Optimal network sizes for most robust Turing patterns}

\author[1,2,3]{Hazlam S. Ahmad Shaberi\thanks{These authors contributed equally to this work.}}
\author[1,2]{Aibek Kappassov\thanks{These authors contributed equally to this work.}}
\author[1,2]{Antonio Matas-Gil}
\author[1,2]{\\Robert G. Endres}

\affil[1]{Department of Life Sciences, Imperial College, London SW7 2AZ, United Kingdom}
\affil[2]{Center for Integrative Systems Biology and Bioinformatics, Imperial College, London SW7 2AZ, United Kingdom}
\affil[3]{Institute of Systems Biology, National University of Malaysia}

\begin{document}
\maketitle

\begin{abstract}
Many cellular patterns exhibit a reaction-diffusion component, suggesting that Turing instability may contribute to pattern formation. However, biological gene-regulatory pathways are more complex than simple Turing activator-inhibitor models and generally do not require fine-tuning of parameters as dictated by the Turing conditions. To address these issues, we employ random matrix theory to analyze the Jacobian matrices of larger networks with robust statistical properties. Our analysis reveals that Turing patterns are more likely to occur by chance than previously thought and that the most robust Turing networks have an optimal size, consisting of only a handful of molecular species, thus significantly increasing their identifiability in biological systems. Broadly speaking, this optimal size emerges from a trade-off between the highest stability in small networks and the greatest instability with diffusion in large networks. Furthermore, we find that with multiple immobile nodes, differential diffusion ceases to be important for Turing patterns. Our findings may inform future synthetic biology approaches and provide insights into bridging the gap to complex developmental pathways.
\end{abstract}

\section*{Introduction}\label{sec:into}

Spatial patterns and structures are common in biological systems, ranging from microbial communities and biofilms to developmental biology and ecological systems \cite{maini_bones_1997} \cite{urban_evolutionary_2020} \cite{martinez-garcia_spatial_2022}.  {In a seminal work, Alan Turing proposed a self-organizing, emergent mechanism for pattern formation based on a diffusion-driven instability \cite{turing_chemical_1952}, later formalized by Gierer and Meinhardt in terms of a slowly diffusing activator and a fast-diffusing inhibitor molecule \cite{gierer_theory_1972}.} According to Turing's definition, the chemical reactions are stable without diffusion, leading to a homogeneous steady state, but become unstable with diffusion, forming a periodic pattern at certain wave numbers. The mathematical analysis is generally based on reaction-diffusion models, implemented using partial differential equations, and linear stability analysis to investigate the effect of perturbations on a steady state \cite{murray_mathematical_2002}. Nonetheless, employing Turing models to understand biological patterns is still a topic of debate.

The Turing mechanism clearly has biological relevance in cell and developmental biology, likely explaining aspects of digit formation in mice, scale-patterns in zebrafish, fingerprint formation, and cortical folds of the fetal brain \cite{lefevre_reaction-diffusion_2010}\cite{sheth_hox_2012}\cite{watanabe_is_2015}\cite{glover_developmental_2023}. However, this mechanism  has two main drawbacks: extreme simplicity and a need for fine-tuning.  {The latter requires parameters to be carefully chosen to produce patterns.} In contrast to simple activator-inhibitor Turing models, biological gene regulatory networks in embryonic development generally consist of hundreds of molecular species and are notable for their immense complexity, hierarchical structure, and tolerance to noise \cite{samarasinghe_comprehensive_2023}. In contrast, the Turing conditions lead to a lack of structural robustness, which is at odds with the noise tolerance and evolutionary adaptability required for such patterning solutions to occur \cite{maini_turings_2012}\cite{vittadello_turing_2021}\cite{krause_turing_2024}. However, there are also bottom-up approaches to further elucidate the issues with Turing models.

As developmental biology is complex, synthetic biology provides an alternative route, following Feynman's mantra: “What I cannot create, I do not understand” \cite{way_what_2017}. By implementing the Turing mechanism from scratch in cells that communicate via chemicals, our understanding of how to generate stable pattern can be systematically improved. Two previous synthetic biology attempts failed due to the lack of differential diffusion, leading to irregular patterns. First, stochastic Turing patterns were engineered in {\it E. coli} cells. The implemented circuit contained self-activation and lateral inhibition, with two diffusible quorum-sensing signals \cite{karig_stochastic_2018}. Second, solitary patterns were engineered in HEK cells using the Nodal–Lefty system \cite{sekine_synthetic_2018}. The issue is that small molecules have roughly the same diffusion constant, although extra nonspecific binding can help slow down diffusion. In a recent work, a more robust three-node network was implemented using synthetic circuits of six genes with small diffusible quorum-sensing molecules and extra control cassettes \cite{tica_three-node_2023}. This showed regular patterns in growing bacterial colonies, but the variability is immense, and robustness is still limited.  {Methods to increase the reproducibility of these patterns, such as solving the inverse problem  given an experimental pattern, have also been recently developed\cite{matasgil_2024_unraveling}.} Hence, while there is progress in the rational design of circuits with specific properties, the lack of control over robustness represents a significant downside in tissue engineering, patterned biomaterial deposition, and bridging developmental programs \cite{cao_programmable_2017}\cite{din_interfacing_2020}\cite{davies_using_2017}.

The need to develop theoretical frameworks beyond the two-equation Turing model has been recognized previously. An exhaustive exploration of all two- and three-node topologies showed that three nodes are, on average, more robust than two nodes, pointing toward complexity increasing robustness \cite{scholes_comprehensive_2019}.  {Note that "nodes" refers here to genes and links to regulatory interactions. These networks should not be confused with Turing models on graphs, which deal with discrete topologies, not continuous space \cite{othmer_instability_1971} \cite{nakao_turing_2010} \cite{kouvaris_pattern_2015} \cite{muolo_turing_2024}.}  More specifically, approximately 60\% of all topologies produced patterns for some parameter combinations, but the parameter space producing Turing patterns was overall minuscule, around 0.1\%. This finding is supported by another study that explored networks with up to four nodes \cite{marcon_high-throughput_2016}. Motivated by Robert May's work in theoretical ecology \cite{may_will_1972} (based on earlier work by Eugene Wigner \cite{allesina_stabilitycomplexity_2015}) a random matrix approach for $N \leq 6$ showed that larger networks add robustness by reducing the "diffusive threshold", i.e., softening the requirement of differential diffusion \cite{haas_turings_2021}. This raises the question: why not investigate even larger matrices? While the diagonalization of Jacobian matrices is a slow $O(N^3)$ process, it is still efficient for significantly larger networks with modern computers. The advantages of exploring random Jacobian matrices are apparent; such an approach produces excellent statistics, and biological realism can be introduced through the distributions and topologies.

Here, we go beyond small Turing networks and use large random Jacobian matrices to systematically explore how network size affects robustness  {of patterning continuous space}. We begin by motivating distributions for Jacobian matrix elements from explicit small Hill-function-based models for gene regulation, as relevant to synthetic biology \cite{tica_three-node_2023}. We then continue by exploring large random networks with up to $N=100$ nodes using corresponding random Jacobian matrices with two diffusers and variable sparsity. We identify an optimal network size $N_\text{opt}\sim5-8$, arising from a tradeoff between stability without diffusion and instability with diffusion, each with opposite $N$-dependencies. This increased robustness relieves the constraints on parameters, known as Turing conditions, including differential diffusion for the two diffusing species. Our results are expected to renew the quest for robust Turing patterns in synthetic biology and to identify Turing modules in larger networks of developmental biology.

\begin{figure}[t]
\centering
\includegraphics[scale=0.35]{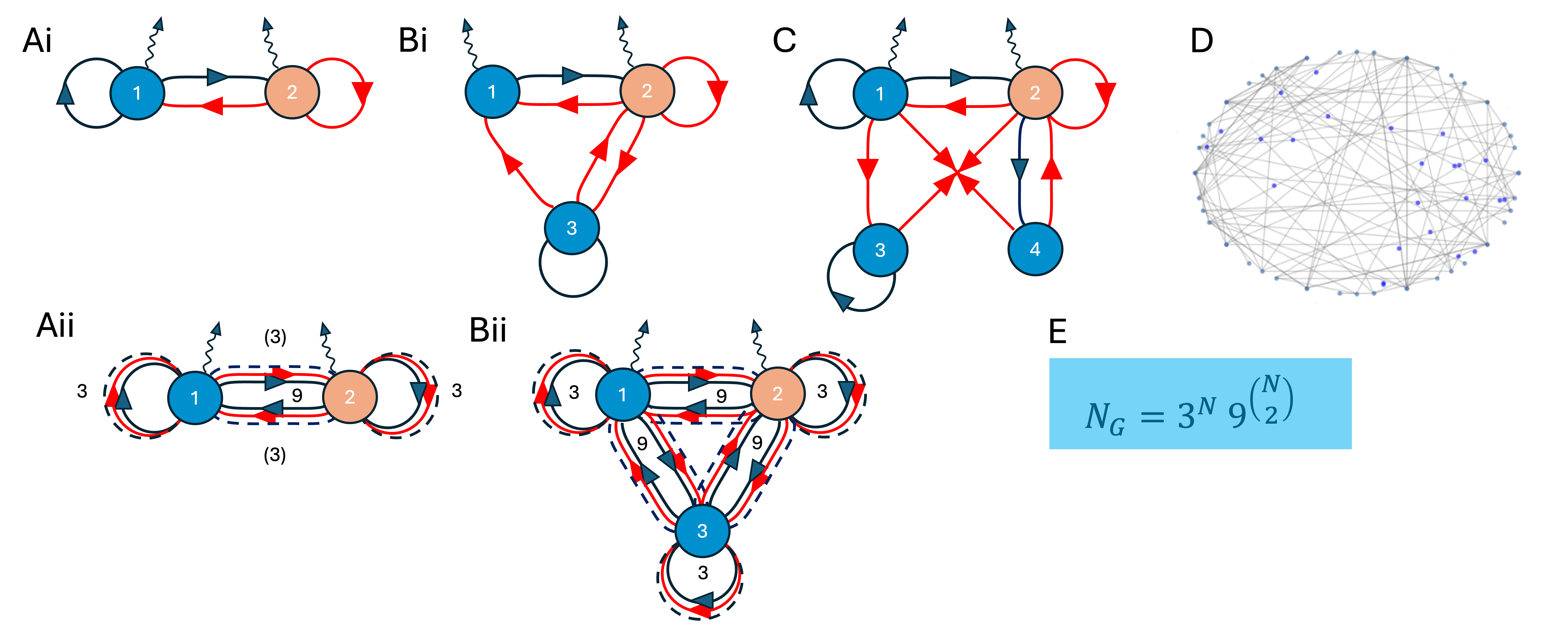}
\caption{{\bf Network graph representations of pre-defined Turing topologies.} (Ai) 2-node Gierer-Meinhardt network. (Bi) 3-node network based on \cite{scholes_comprehensive_2019, tica_three-node_2023}. (C) Extended 4-node network. (D) Large $N>\!\!>1$ network. (E) Number of different Turing-network topologies for $N$ nodes, exemplified by all 2 and 3-node networks in (Aii) and (Bii), respectively. Dark blue (red) arrows indicate activation (inhibition), dashed lines denote missing links, and wiggly lines indicate diffusion.}
\label{fig:fig1}
\end{figure}

\section*{Results}\label{sec:results}
We are interested in the robustness of large Turing networks, that is reaction-diffusion models with many molecular species with a large parameter space to support diffusion-driven instabilities. Fig. \ref{fig:fig1} shows example networks, ranging from small ($N=2$ nodes) to large ($N>\!\!>1$). We restrict ourselves to two diffusing species similar to Turing’s original paper. In order to understand how to model large networks we begin by investigating examples of smaller networks to gain intuition. For a network of $N$ nodes and hence $N$ species with time and space dependent concentrations $\bm{X}(\bm{r},t)=\{x_1(\bm{r}, t), x_2(\bm{r}, t), \dots, x_N(\bm{r},t)\}$, the general dynamics are given by the following partial differential equations
\begin{equation}
    \frac{\partial x_i}{\partial t}  =  f_i(\bm{X}; \bm{\theta}) + D_i \nabla^2 x_i,\label{eq:dXdt}
\end{equation}
where $\bm{D}=diag(D_1, D_2, 0, \dots, 0)$ is the diagonal diffusion matrix, and Laplacian $\nabla^2=\sum_{i=1}^d\partial^2/\partial r_i^2$ in $d$ dimensions, given by the sum of second spatial derivatives. For simplicity, we assume an infinite domain, allowing us to neglect boundary effects and to use a continuous wave number.  {The nonlinearities are given by the interactions between the nodes. As shown in Fig. \ref{fig:fig1}Aii and Bii, we consider all possibilities for each connection, depicted as black for activation, red for inhibition and dotted for no interaction. We will refer to activation and inhibition nodes as positive and negative edges respectively. Curved lines denote diffusion, which as mentioned will only be considered for the first two nodes}. As our motivation is to learn how to build these with synthetic gene circuits, we model the activating and inhibiting interactions with Hill functions, together with basal expression and degradation terms:

\begin{equation}
\label{eq:f_i}
    f_i(\bm{X}; \bm{\theta}) = b_i + V_i\prod_{j \in S_j^+} \frac{1}{1+\left(\frac{K_{ji}}{x_j}\right)^{n_{ij}}}
    \prod_{j \in S_j^-}\frac{1}{1+\left(\frac{x_{j}}{K_{ji}}\right)^{n_{ij}}} - \mu_i x_i,
\end{equation}
with $\bm{\theta} =\{\bm{b}, \bm{V}, \bm{K}, \bm{n}, \bm{\mu}\}$ respectively the sets of basal and maximal expression rates, concentration thresholds, Hill coefficients, and degradation rates. Furthermore, $S_i^{+ (-)}$ denotes the set of positive (negative) edges ending in node $i$, where the different regulators act (and saturate) independently (non-competitively) of each other. Following \cite{scholes_comprehensive_2019}, we sample parameters from a wide range of allowed values in arbitrary units. 

To understand whether such parameter combinations produce Turing patterns, we use linear stability analysis following Turing’s approach (see Methods for additional details). Briefly, first the model needs to have a stable homogeneous steady state $\bm{X}^*$ without diffusion, defined by
\begin{equation}
    \bm{f}(\bm{X}^*; \bm{\theta}) = 0,
\end{equation}
which we solve with the Newton-Raphson method using different initial conditions. Second, we linearize the dynamic equations assuming small perturbations around steady state, using $x_i(\bm{r},t)=x_i^*+\delta x(\bm{r},t)$, leading to 
\begin{equation}
    \frac{\partial ({\delta \bm{X}})}{\partial t}  =  \bm{J}\delta\bm{X} + \bm{D}\nabla^2 \delta\bm{X},\label{eq:ddXdt}
\end{equation}
with Jacobian matrix $\bm{J}$ of first derivative with matrix elements $J_{ij}=\partial f_i/\partial x_j$.

\begin{figure}[t]
\centering
\includegraphics[width=\textwidth]{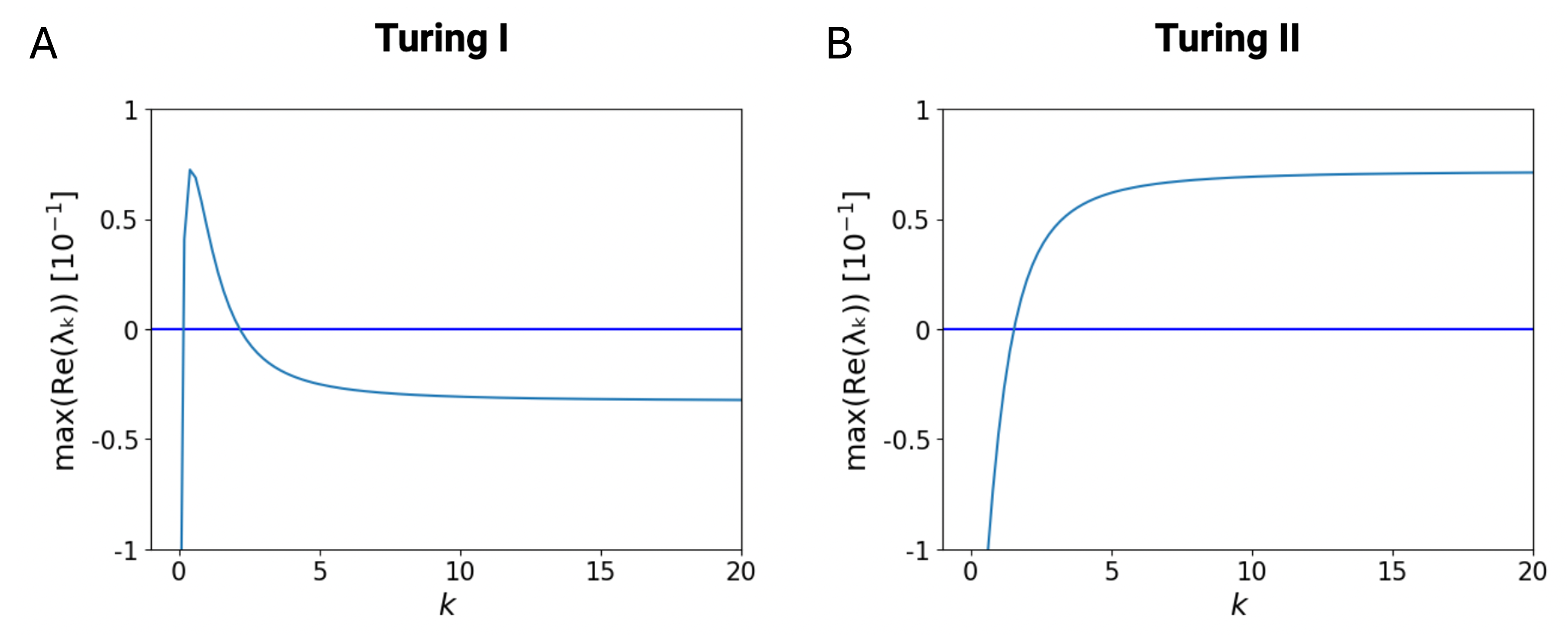}
\caption{{\bf Dispersion relations illustrating different Turing instabilities.} (A) Turing type I instability with a peak and negative real part for $k>\!\!>1$. (B) Turing type II instability which stays positive (note this is however not possible when all species diffuse). The $y$ axis represents the maximum real part of the eigenvalue and the $x$ axis shows wave number $k$. The plots are generated from random matrices with network size $N = 10$ and variance $\sigma^2 = 0.25$.} 
\label{fig:fig2}
\end{figure}

To remove the second derivatives, we Fourier transform in the space domain, or more simply apply a wave-like perturbation $\delta x(\bm{r},t)=\delta\tilde x(t)\exp{(i\bm{k}\bm{r})}$ with $\bm{k}=\{k_1, k_2, \dots, k_d)$ and $\bm{k}\bm{r}$ the scalar product between the $\bm{k}$ and $\bm{r}$ vectors. Plugged into Eq. \ref{eq:ddXdt}, this leads to a new Jacobian matrix with modified diagonal matrix elements and extra parameter $k=|\bm{k}|$, given by
\begin{equation} \label{eq:jacobian_with_diff}
\begin{split}
 \boldsymbol{J}(k) & = \left. \boldsymbol{J}_0 \right|_{\boldsymbol{X}^*} - k^2 \bm{D},
\end{split}
\end{equation}
indicating that for a rotationally invariant system, the dependence is only on the modulus of $\bm{k}$, rendering linear stability analysis an effective one-dimensional problem. As a result, diagonalization of the modified Jacobian matrix leads to $k$-dependent eigenvalues, called dispersion relations, where the one with the largest real part, termed $Re(\lambda_\text{max}(k))$, determines the stability. For a Turing instability, we require the homogeneous steady state without diffusion to be stable (corresponding to $k$=0), i.e. with a negative real part given by $Re(\lambda_\text{max}(k(0)))<0$. Furthermore, with diffusion, we require instability for a wave number ($k>0$) and thus a positive real part given by $Re(\lambda_\text{max}(k))>0$. For a classic Turing instability leading to a well-defined wave pattern, there is a $k_\text{max}>0$, for which the dispersion has a clear maximum, and for $k\rightarrow\infty$, the system becomes stable again (negative real part of eigenvalue). This type of instability is called Turing I (Fig. \ref{fig:fig2}A). However, there are other possibilities, such as the instability for $k\rightarrow\infty$ (Turing II in Fig. \ref{fig:fig2}B).  {In the latter case, there is no well defined wavelength for pattern formation, and hence Turing II cases are generally not considered in the following.} (Note that this is a simplified classification scheme - in other schemes our Turing I is given by Turing Ia, and our Turing II is given by Turing Ib, IIa, and IIb with subtle differences \cite{scholes_comprehensive_2019}.) Yet, another type of instability is the Turing-Hopf, for which the imaginary part can additionally lead to oscillations (not shown).

\begin{figure}[t]
\centering
\includegraphics[width=\textwidth]{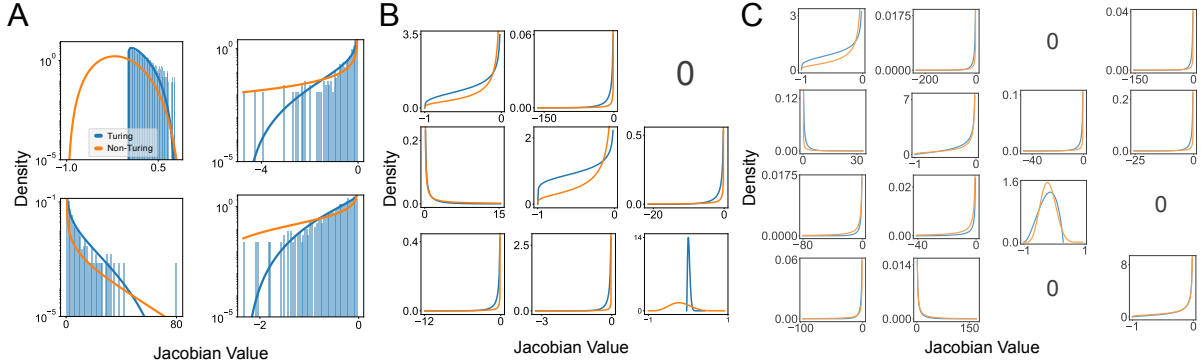}
\caption{{\bf Empirical distributions of Jacobian matrix elements.} (A) Histograms for parameters leading to Turing I instabilities (light blue) and fits to the beta distribution for Turing I (blue line) and non-Turing (orange line). Empirical distributions are computed based on parameter sampling for the pre-defined 2-4-node network topologies in Fig. \ref{fig:fig1}Ai, Bi, and C for 2-node (A), 3-node (B), and 4-node (C) networks. Note some matrix elements are zero due to missing links in the network (or entries in the adjacency matrix).}
\label{fig:fig3}
\end{figure}

Focusing initially on small networks ($N = 2-4$), we employed Latin hypercube sampling of $10^7$ parameter sets and filtered them for Turing I instabilities. Fig. \ref{fig:fig3} shows the fits of the empirical histograms of the Jacobian matrix elements ($j$) to the beta distribution $B(j; \alpha,\beta)$ (see Methods for details on parameter searches and fitting procedure). The beta distribution depends on two parameters, $\alpha$ and $\beta$, allowing a wide range of distributions from Gaussian-like symmetric to non-Gaussian shapes with various levels of skewness and kurtosis. We generally observe that the Turing-I matrix elements are more narrowly distributed in line with Turing I having to fulfill Turing conditions, leading to a subset of parameters and hence Jacobian matrix elements. This can be clearly seen in Fig. \ref{fig:fig3}A (see Fig. S1 in the Supplementary Figures for additional histograms). To fulfill the Turing conditions, the (1,1) Jacobian matrix elements need to be positive for activation, which is the case for the histogram and fitted line (blue), but not for the orange line, representing the non-Turing case.  {Note, while Node 1 is an activator, degradation contributes to the diagonal matrix elements, potentially leading to negative values.} Furthermore, the (2,2) matrix elements need to be negative for inhibition, which is here already fulfilled for non-Turing due to model constraints (degradation). In contrast, the (1,2) and (2,1) matrix elements can, respectively, be either negative and positive (here the case even for non-Turing due to model constraints), or the other way around (however not for our 2-equation model) \cite{murray_mathematical_2002}. We generally did not observe multimodal distributions, which indicates a certain level of simplicity in the distributions.

\begin{figure}[t]
\centering
\includegraphics[width=13cm]{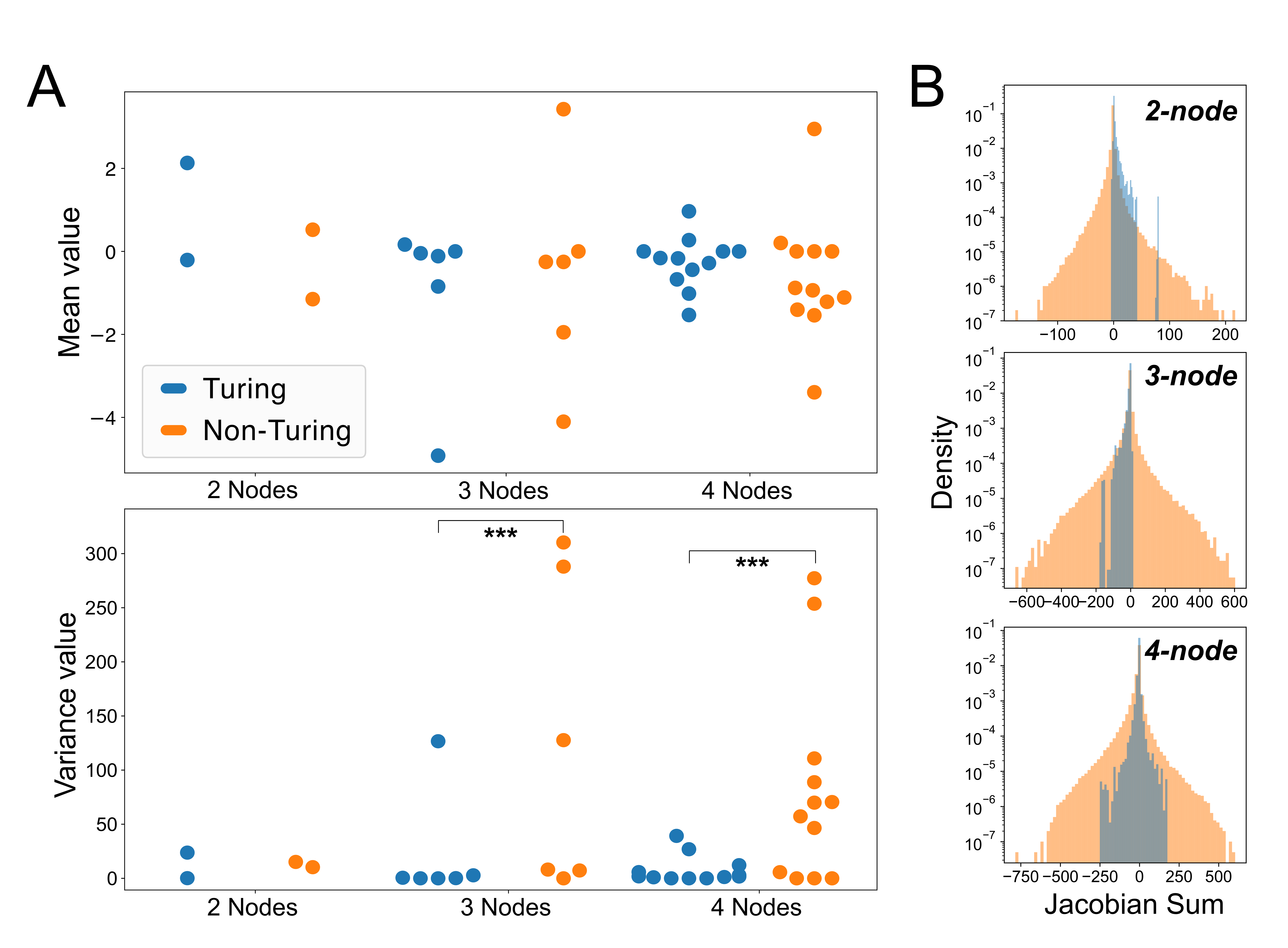}
\caption{{\bf Jacobian matrix elements are different for Turing I and non-Turing.} (A) Beehive plots of the mean (top) and variance (bottom) of the empirical distributions of the non-zero off-diagonal Jacobian matrix elements from Fig. \ref{fig:fig3} (based on the pre-defined 2-4-node network topologies in Fig. \ref{fig:fig1}Ai, Bi, and C). A statistical Brown-Forsythe test was utilized to verify whether the difference in variance between each pair of Turing and non-Turing distributions was statistically significant. For all pairs of the Jacobian distributions except one, we found that the test showed a statistically significant difference between the variances. Furthermore, 12 out of 16 $p$-values were under 0.001, showing a strong statistical significance. (B) Sum of off-diagnal matrix elements, showing a clear visual difference between Turing I (light blue) and non-Turing (light orange) for the 2-4 node topologies. The distributions become increasingly symmetric, and ultimately Gaussian, for large numbers of nodes.} 
\label{fig:fig4}
\end{figure}

The difference between Turing and non-Turing Jacobian matrix elements is further visible in Fig. \ref{fig:fig4}, where panel A shows beehive plots based on the empirical data from Fig. \ref{fig:fig3}. Clearly, the variances of the off-diagonal matrix elements of the non-Turing cases are much broader than the Turing-I cases. This is further illustrated in panel B, showing the sum of the Jacobian matrix elements (equivalent plots can be made for the mean of those random variables). For larger matrices, we expect to observe the central limit theorem, with the sums calculated from $N(N-1)$ off-diagonal Jacobian matrix elements. For increasing matrix size,  {this trend is already evident for $N=4$, although there is still a significant cusp at zero}: The mean is approximately zero, the variance approaches a finite value (as variances are positive numbers), and the skewness and kurtosis vanish. Hence, not surprisingly, on average a Gaussian distribution with zero mean and finite variance describes the sum of Jacobian matrix elements. Note that this trend would be even stronger, if we average over all possible network topologies with $N$ nodes, which scales as $N_G = 3^N 9^{\binom{N}{2}}$ and hence is superexponentially increasing with $N$ (see Fig. \ref{fig:fig1}E). In the following, we exploit these insights and use random Jacobian matrices directly.

Inspired by Robert May‘s statistical treatment of non-equilibrium ecological communities that studied asymmetric random Jacobians to represent linearized population dynamics, contrasting Wigner's symmetric (Hermitian) random matrices \cite{may_will_1972}\cite{stone_feasibility_2018}, we analyzed Turing instabilities by sampling asymmetric random Jacobian matrices that represent the linearized reaction dynamics. Similar to May's "neutral interaction" model, our Jacobian matrix without diffusion is given by
\begin{equation} \label{eq:random_matrix}
\bm{J_0} = \bm{G} - \bm{I}, 
\end{equation}
where $\bm{G}$ is a random matrix with elements $g_{ij}$ randomly assigned from a Gaussian distribution that has mean value zero and variance Var$(g_{ij}) = \sigma^2$, but with diagonal terms $g_{ii} = 0$. For simplicity (and  convenience), the latter are modeled by the identity matrix, $\bm{I}$, responsible for providing stability. In such a Turing system, a positive value of $J_{ij}$ (or $g_{ii}$) implies activation while a negative value implies inhibition, similarly to previous matrix representations of Turing systems \cite{scholes_comprehensive_2019}\cite{diego_key_2018}. The identity matrix $\bm{I}$ represents the degradation of the species. Further self-interactions are neglected, leading to the curious result that there are no Turing instabilities for $N=2$ (see proof in Methods). While the Gaussian distribution is motivated by Fig. \ref{fig:fig4}, there is no reason to expect the reaction kinetics in a reaction-diffusion system to be independent as there was no reason for the population dynamics in the work of May \cite{may_will_1972} to be independent.  {However, detailed understanding of realistic models and their parameters, or Jacobian matrices of linearized models is missing. Hence, the statistical random-matrix approach has been powerful in understanding the principles of stability in population dynamics \cite{allesina_stability_2012,servan_2018} and small Turing systems ($N\leq 6$) \cite{haas_turings_2021}.}

\begin{figure}[t]
\centering
\includegraphics[width=\textwidth]{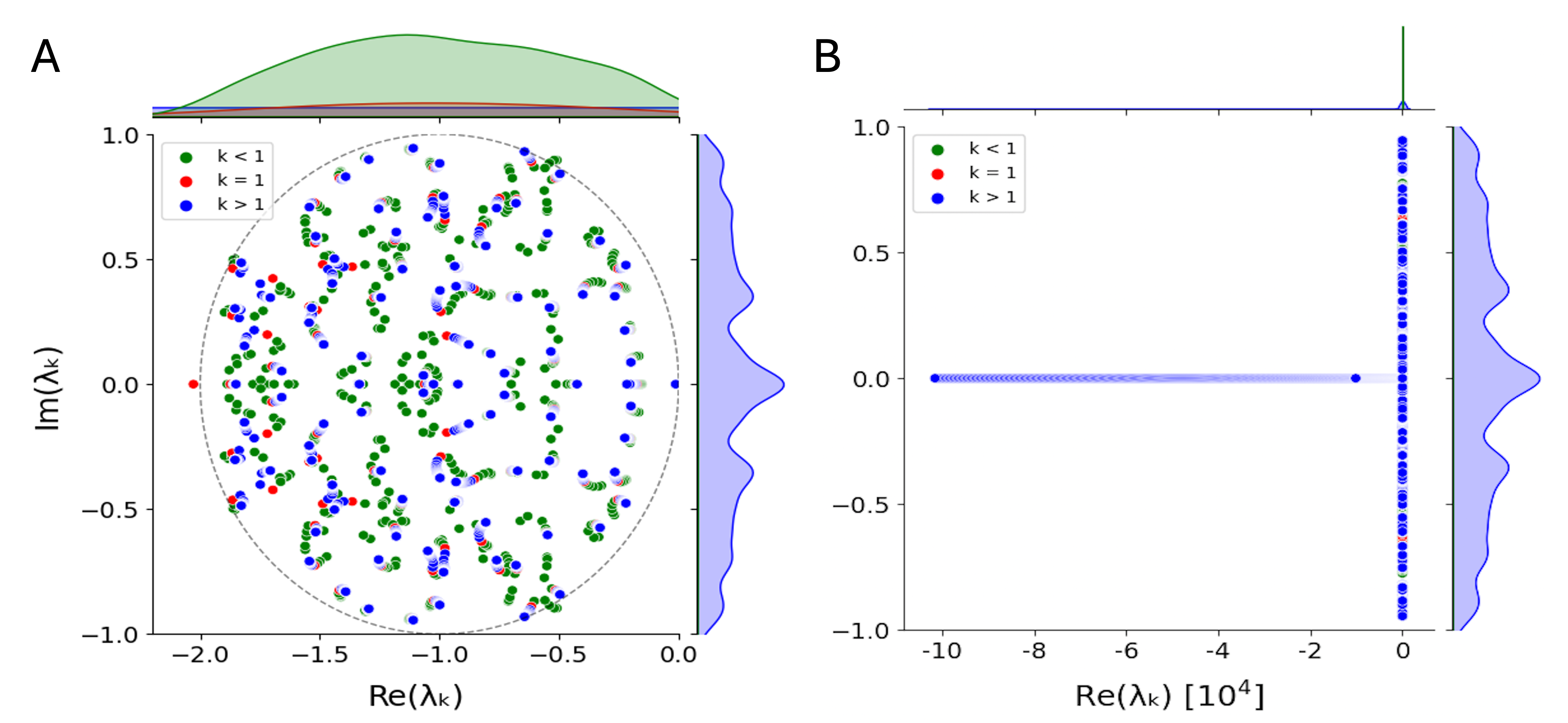}
\caption{{\bf Eigenvalue distribution of a large random network.} (A) Circular scatter plot and corresponding kernel density estimations of eigenvalues in the complex plain  {(green at the top and blue on the right)} for network size $N = 100$, variance $\sigma^2 = 0.01$, and diffusion constants $D_1 = 1$, $D_2 = 10$.  {$k$-dependent movement of the eigenvalues is visible, and finite-$N$ effects in the kernel density estimates as well, such as suppression of the density at the boundary of the circle.} While the system without diffusion has all the eigenvalues distribute around $\gamma=1$, the same system with diffusion has most but not all of its eigenvalues distribute in a circle with radius $\gamma=1$. (B) Full eigenvalue distribution  {at a large scale with focus} on the outliers  {(the eigenvalue circle now appears like a vertical line.)} Green, red and blue points correspond to wave numbers $k<1$, $k = 1$ and $k>1$, respectively.}
\label{fig:fig5}
\end{figure}

Next, we investigate the eigenvalue distributions of such random matrices (see Fig. \ref{fig:fig5}A). In the absence of diffusion (i.e. at $k = 0$), the eigenvalues are distributed in a circle with radius $\gamma = \sigma\sqrt{N}$ according to May's "circular law", derived from random ecological communities \cite{may_will_1972}\cite{stone_feasibility_2018} and exact in the asymptotic limit. In the presence of diffusion, the eigenvalue distribution can be described as having "in-bulk" and "outlier" distributions (see Fig. \ref{fig:fig5}B). The outlier distribution has extreme negative real parts, following approximately $-N-k^2\sum_{i=1}^ND_i$
(see Methods), which tends to minus infinity for increasing $k$ and network size $N$. As we are only interested in the maximal real part to determine the linear stability, it is safe to ignore the "outliers"  and only focus on the “in-bulk” eigenvalue distribution, containing the maximal real eigenvalues.

\begin{figure}[t]
\centering
\includegraphics[width=\textwidth]{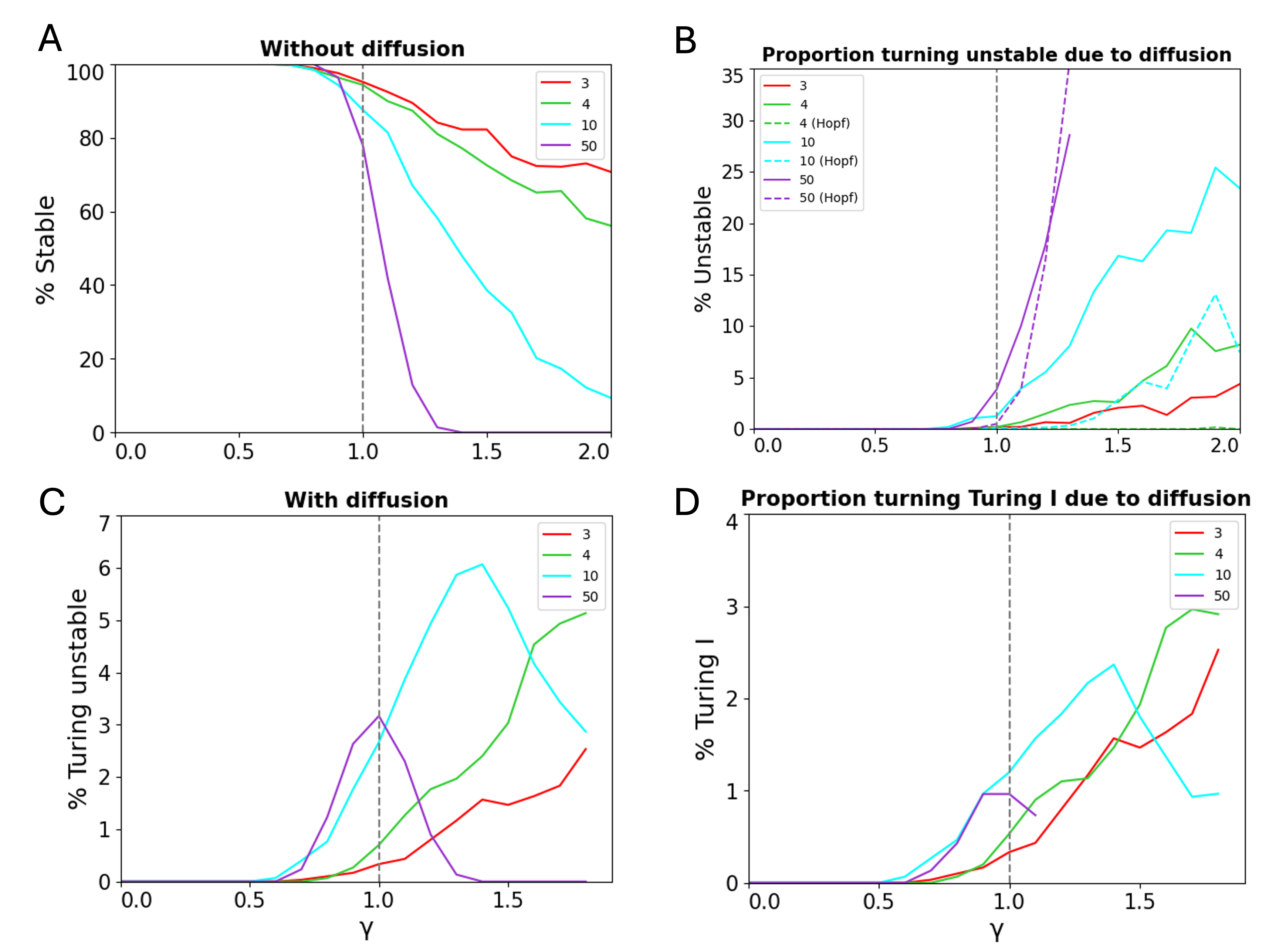}
\caption{{\bf Relationships between network size $N$ and radius $\gamma=\sigma\sqrt{N}$.} $10^3$ random matrices are sampled for each $N$ and $\gamma$ pair. (A) Percentage of stable random matrices without diffusion. (B) Percentage of previously stable random matrices turning unstable with diffusion. 'Hopf' indicates Turing-Hopf. (C) Percentage of random matrices leading to Turing instabilities (Turing I, II, and Turing-Hopf). (D) Percentage of random matrices leading to Turing I instabilities. The key point is $\gamma=1$, while for larger $\gamma$ values the statistics become unreliable.}
\label{fig:fig6}
\end{figure}

We subsequently sampled thousands of random matrices to analyse how network size affects the stability of a reaction-diffusion system. We observed three relationships between network size and stability. Firstly, in the absence of diffusion, a random reaction-diffusion system with fixed radius $\gamma$ is likely more stable for smaller network size (Fig. \ref{fig:fig6}A). 
Asymptotically, the circular law predicts a step function for $N\rightarrow\infty$, i.e. stability for radii $\gamma<1$ (as circles are fully to the left of the imaginary axis) and instability for $\gamma>1$ (as circles cross the imaginary axis) \cite{allesina_stabilitycomplexity_2015}.  
Secondly, a previously stable system without diffusion is more likely to turn unstable (i.e. having a diffusion-driven instability) when diffusion is present for larger network size (Fig. \ref{fig:fig6}B). 
 {Therefore, thirdly, out of all random matrices, Turing instabilities are maximized (i.e. found the most) at an optimal intermediate network size (Fig. \ref{fig:fig6}C).}

As the Turing-Hopf instability is competing with the Turing instability, and Turing I is of importance in producing spatial patterns, we analyzed the effects of network size on Turing I and Turing-Hopf instabilities. Out of stable systems without diffusion, the systems are likely to become Turing-Hopf unstable in the presence of diffusion with larger network size (Fig. \ref{fig:fig6}B). This trend for the Turing-Hopf instability is similar to the trend for Turing I. However, particularly at a very large network size, the Turing-Hopf instability becomes more likely to occur and ultimately out-competes Turing I, as can be seen for $N = 50$ in the figure. A similar trend is observed for Turing I relative to the overall Turing instabilities of all types: Turing I is more likely for larger network size being approximately half the occurrence of all Turing instabilities. Among all Turing instability types, we found that Turing II is the most common to arise (4.2\% maximum occurrence of all random reaction-diffusion systems), followed by Turing I (3.6\%).

\begin{figure}[t]
\centering
\includegraphics[width=\textwidth]{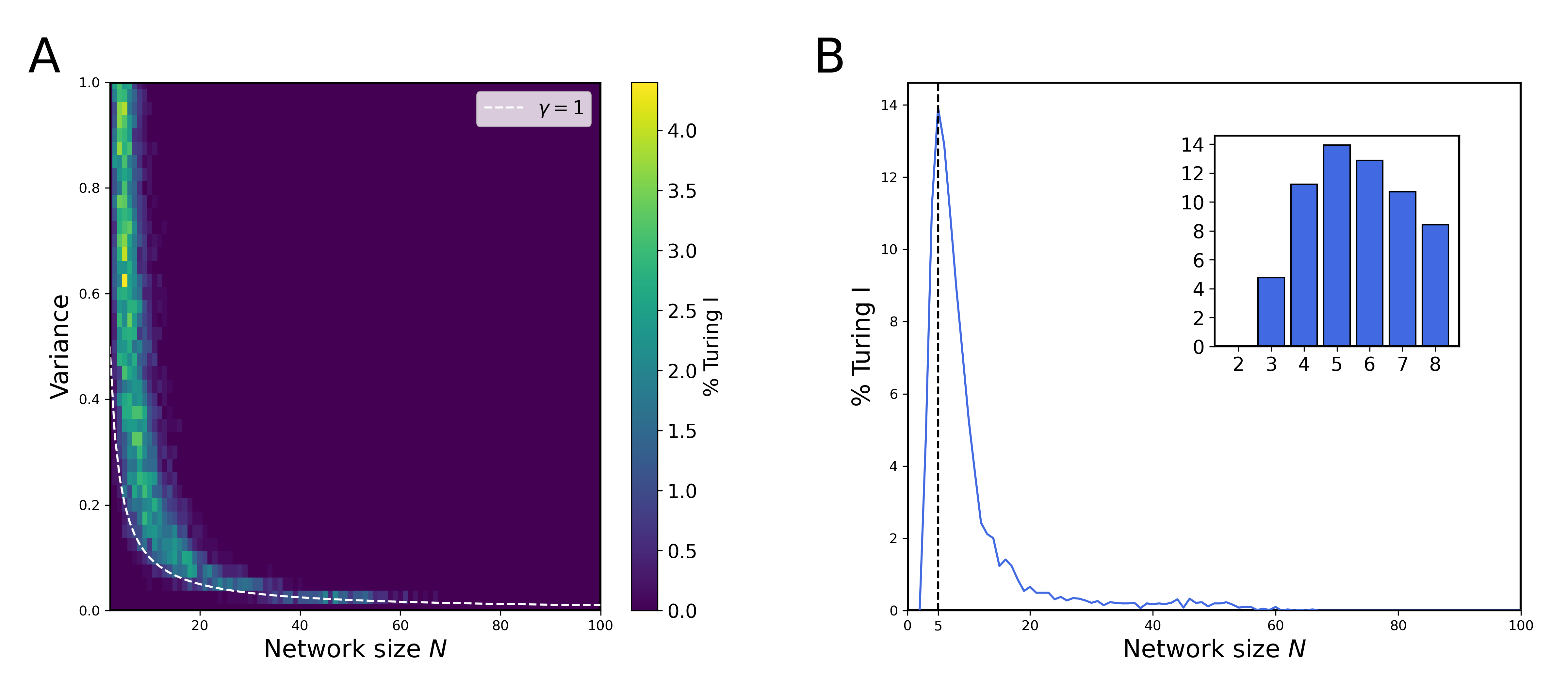}
\caption{{\bf Optimal network size for highest Turing robustness.} (A) Heat map of percentage occurrence of Turing I in random matrices of different network sizes $N$ and variances $\sigma^2$ for $\gamma=1$ and zero off-diagonal element sparsity (or connectivity $C = 1$). (B) Corresponding percentage shares of each network size $N$. For our parameters, the optimal network size is $N = 5$.  {The inset provides a more detailed view of network sizes from 2 to 8 for finer resolution.}}
\label{fig:fig7}
\end{figure}

Without having an experimental understanding of the variability of interaction strength between nodes, an optimal network size should produce Turing I instabilities for a wide range of variances. Such a network size would be most robust to changes in interaction strength, which is a desirable outcome from a biological perspective. To find the exact network size that is the most optimal, we plotted the occurrence of Turing instabilities for different network sizes and variances from $N=2$ to $100$ (Fig. \ref{fig:fig7}A). Based on the heat map, the majority of the Turing instabilities arose just above $\gamma=1$, as we should expect since $\gamma>\!\!>1$ would likely result in an unstable system without diffusion, but $\gamma<\!\!<1$ would make it harder for a stable system to become unstable with diffusion. 
In fact, the occurrence of Turing is "pinned" to radius $\gamma=1$ for large $N$, and the percentage density then follows closely the circular law $\sigma^2\sim1/N$. 

Is there an optimal network size for highest robustness of the occurance of Turing patterns? We expect such an $N_\text{opt}$ as there are no Turing patterns for $N=2$ and for $N\rightarrow\infty$. Turing patterns disappear as the eigenvalue distribution becomes a step function, allowing no instability with diffusion. To confirm our intuition, we then calculated the percentage of Turing I for each network size from the heat map (Fig. \ref{fig:fig7}B), demonstrating that $N_\text{opt}=5$ is the most optimal network size based on its highest percentage of Turing patterns. Thus, for a network with two diffusible species, having additional three immobile nodes gives the optimal network size for Turing instabilities. Compared to $N = 3$ with one non-diffusible species (4.73\%), Turing instabilities are three times more likely to occur for $N+\text{opt} = 5$ (13.86\%).

 {The peak is however not at the first allowed $N$ value, i.e. $3$, but at a higher value, pointing towards another principle for this maximum in Turing frequency. For finite-\(N\) random matrices and \(\gamma \gtrsim 1\), the scaling of the eigenvalue density, projected onto the real axis, near \(x = 0\) can be approximated as
   \[
   \rho(0) \sim \frac{1}{\pi} \sqrt{\gamma^2 - 1} \cdot \left(1 - \frac{1}{2 \gamma^2 N} + \mathcal{O}\left(\frac{1}{N^2}\right)\right)
   \]
(see the Supplementary Text for details). For small \(N\), the density of marginal stable matrices i.e. near $x=0$, is suppressed. This is due to eigenvalues being correlated and biased toward the center of the eigenvalue circle due to finite $N$ effects. As a result, there are less Turing cases for small $N$ than expected based on the asymptotic form of Robert May's circular law. This ultimately leads to the value $N_\text{opt}=5$.}

While our random matrices are fully connected, it was suggested that biological networks tend to be sparsely connected due to its evolutionary advantage in preserving robustness \cite{pavlopoulos_using_2011}. Let us assume that the probability of observing $k$ connections is binomially distributed $\binom{n}{k}p^k(1-p)^{n-k}$ where $n$ the number of matrix elements to consider ('trials') such as $n=N(N-1)$ for the number of off-diagonal matrix elements, and $p$ the probability for having a connection ('success'). As the variance $np(1-p)$ is largest for $p=0.5$, corresponding to the maximum entropy distribution, we would expect on average 50\% connections. This corresponds to the maximum number of possible network structures found by evolution. However, does sparsity also translate into higher Turing robustness? We incorporated sparsity but found that its effect is minor, without qualitatively changing our results (see Figs. S2 and S3 in the Supplementary Figures). The peak value of the Turing I percentage slightly reduces with sparsity ($\approx 14$, $12$, and $10$\% for respective sparsities  $0$, $25$, and $50$\%), while the corresponding optimal network sizes increase slightly ($N_\text{opt}=5$, $7$, and $8$ for above sparsities). Hence, overall the robustness of Turing is reduced with sparsity, pointing to feedback being important for pattern formation. Note the circular law also applies to sparsity, but needs modification. The radius is just rescaled to $\gamma=\sigma\sqrt{NC}$ where connectivity $C=p$ is the probability for having an off-diagonal matrix element \cite{allesina_stabilitycomplexity_2015}. This points to our results being broadly representative for a large range of random Turing networks.

 {Additionally, we tested full Gaussian matrices with Gaussian-distributed diagonal elements (allowing self-interactions), both with and without self-degradation ($-1$), as well as anti-symmetric matrices with diagonal elements fixed at -1, as in the original setup. The anti-symmetry is designed to capture correlations arising from potential mass-conserved reactions. The full Gaussian matrices with fixed degradation produce an optimal network size (see Fig. S4-S5), whereas the anti-symmetric matrices with $-1$ on the diagonals do not produce Turing patterns at any network size or variance, in line with known theorems \cite{petersen2012matrix}. This is because the eigenvalues of real anti-symmetric matrices are purely imaginary or zero, and adding -1 to the diagonal shifts them along the real axis, leaving no positive real parts to enable Turing instabilities. For the full Gaussian matrices, Turing patterns are possible for $N=2$ when self-interactions are present, as degradation is no longer fixed at $-1$ for all nodes. However, when there is no fixed degradation of $-1$, no clear peak in network size emerges, as the eigenvalue distribution becomes more diffuse without a dominant eigenvalue near $-1$ to stabilize the system.}
   
\begin{figure}[t]
\centering
\includegraphics[width=13cm]{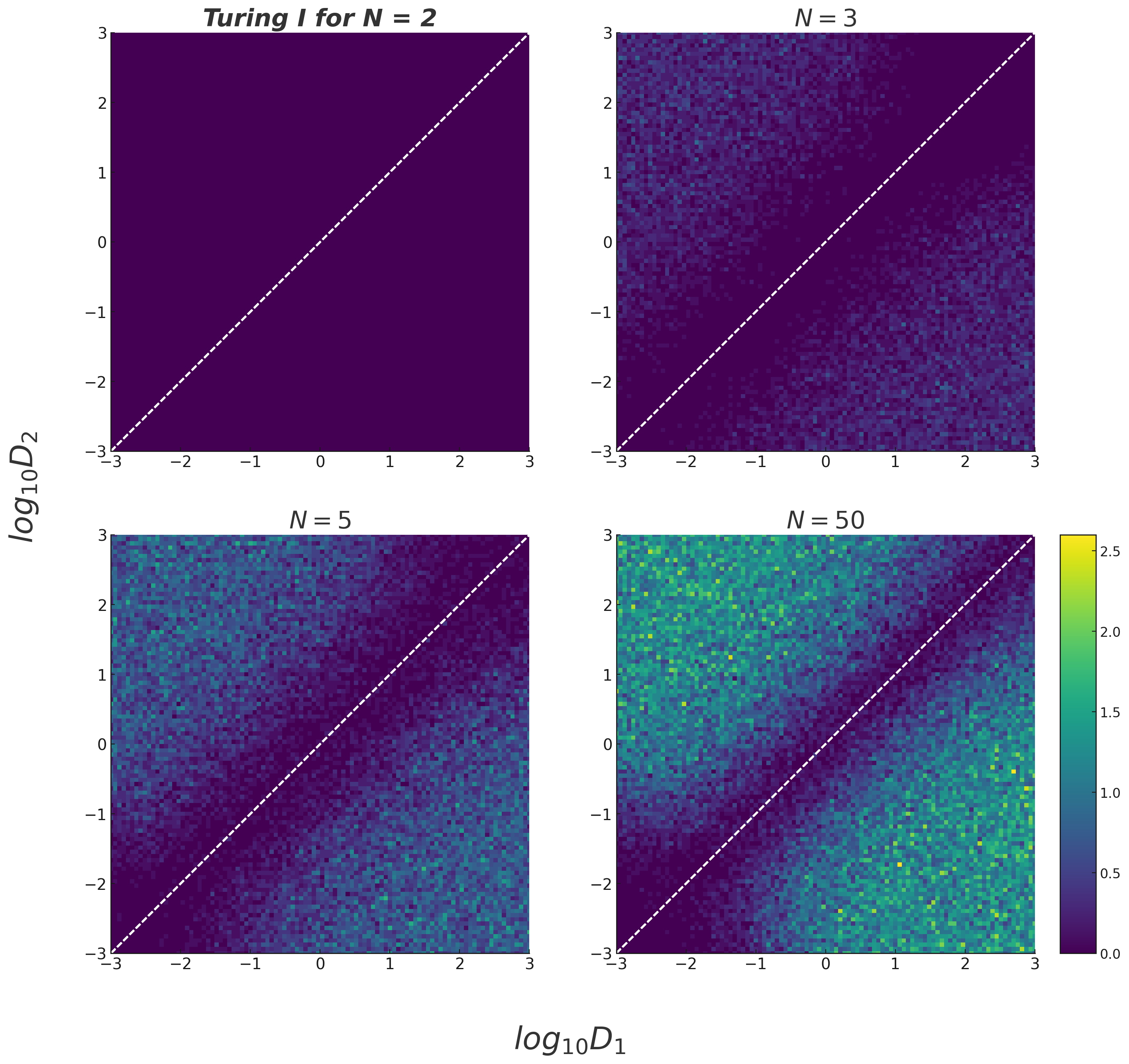}
\caption{{\bf Effect of diffusion constants on Turing pattern formation.} Heat map of percentage occurrence of Turing I in random matrices for different diffusion parameters, $D_1$ and $D_2$: (A) $N = 2$, $\sigma^2 = 0.5$; (B) $N = 3$, $\sigma^2 = 0.33$; (C) $N = 5$, $\sigma^2 = 0.2$  (D) $N = 50$, $\sigma^2 = 0.02$. For increasing $N$, the constraint of differential diffusivity begins to vanish.}
\label{fig:fig8}
\end{figure}

A major constraint on Turing patterns is that the inhibitor has to diffuse much faster than the activator. How does having immobile nodes change this constraint? Previous work on small networks hinted towards a softening of this constraint \cite{marcon_high-throughput_2016,scholes_comprehensive_2019,haas_turings_2021}. By varying the diffusion of the two diffusible species modeled in our random reaction-diffusion systems, we sampled systems of $N = 2$, $3$, $5$, and $50$ nodes, with $N=2$ again not supporting Turing for our random matrices with equal degradation of $-1$. Importantly, we found that differential diffusivity is important for Turing I instabilities to arise for very few immobile nodes, but not for systems with many immobile nodes: Based on Fig. \ref{fig:fig8}, a network of $N = 3$ with one immobile node has nearly zero percent Turing instabilities arising at equal diffusivity (i.e. at the identity line $D_1/D_2=1$). As the diffusion ratio deviates from $1$, Turing instabilities become more likely for $N = 3$, creating a distinct gap around the identity line. Notably, the plot is symmetric with respect to the identity line for all $N$ since the assignment to activator or inhibitor is arbitrary. In contrast, in the presence of more immobile nodes, Turing instabilities can arise at equal diffusivities: As network size increases, the gap at the identity line becomes more narrow, indicating Turing instabilities occurring at a wider combination of the diffusion parameters (see also Fig. S6 in the Supplementary Figures). For the large network size of $N=50$, we observed a similar occurrence of Turing instability for the majority of the diffusion parameter combinations, except for when the diffusion parameters are both extremely low, approximating the case of no diffusion. Since the Turing instability is a diffusion-driven instability, it is reasonable to have no Turing instability arising here. Additionally, there is no 'dark line' (indicating zero or low percentage) at the very bottom or very left of all surface plots, indicating that a Turing instability can arise in systems with only one diffusible species. The absence of differential diffusion is even stronger when consider all Turing cases (Turing I, II, and Turing-Hopf) - see Fig. S7 of the Supplementary Figures. In summary, having numerous immobile nodes significantly relaxes the constraint of differential diffusivity from the classic Turing conditions.

\newpage

\section*{Discussion}

Turing networks have traditionally been small in size, making a direct mapping to complex developmental pathways with many unknown molecular species and parameter values difficult. Another drawback of using Turing models in biology has been the requirement to fine-tune parameters to fulfill Turing's conditions on stability without diffusion and instability for certain wave numbers with diffusion, in contrast to biology's robustness and evolvability \cite{wagner_robustness_2007}. Here, we circumvented these issues by using large random matrices to describe linearized Jacobian systems, greatly extending previous studies on $N \leq 6$ \cite{haas_turings_2021}. By extensively sampling matrix elements, we obtained excellent statistics for networks with up to 100 molecular species.

Applying Robert May's circular law \cite{may_will_1972} to cases without and with diffusion of two designated species, in line with Turing's original model \cite{turing_chemical_1952}, we identified an optimal network size for maximal robustness, reaching 13.86\% of sampled matrices. This value is orders of magnitude higher than previous estimates for small networks \cite{scholes_comprehensive_2019}. Our results are robust even when sparsity is introduced, leading only to minor changes in the observed Turing percentage or optimal network size. Importantly, an optimum in Turing robustness emerges in our model due to the tradeoff between maximizing stability without diffusion, which favors small networks with eigenvalues having negative real parts, and maximizing instability with diffusion, which favors large networks with eigenvalues having positive real parts. According to the circular law, small networks have small radii for the distribution of their eigenvalues, while large networks have large radii. For increasingly large Turing networks, we further found that differential diffusivity is not required, in line with studies on Turing topologies \cite{marcon_high-throughput_2016}. Having many immobile nodes effectively allows for delays and hence different diffusion constants. We hope these results shed new light on Turing mechanisms in developmental systems and help make Turing mechanisms more relevant to quantitative biology. 

 {While we focus on the sweet spot of Turing network size for maximal robustness, a recent study of very large networks supports our findings, that complex random networks with $N>\!\!>1$ do not tend to spontaneously form Turing patterns \cite{piskovsky_will_2024}. When not constraining $\sigma^2$, this is because eigenvalue circles cross over to the positive real axis, leading to instability even without diffusion. Even when constraining the radius to $\gamma=1$, this is still the case, as we found, since the eigenvalue distribution becomes a step function, eliminating the important marginal stable cases. Remedies for overcoming above mentioned issues are identified in \cite{piskovsky_will_2024}, such as allowing for non-diffusive interactions including cross-diffusion, nonlocal interactions, or advection (chemotaxis) \cite{zhao_chemotactic_2023}.}

To make our model interpretable, we used a number of simplifying assumptions, resulting in limitations of our predictions. A major simplification is our diagonal matrix elements, which are set to the same value, representing equal degradation rates for all species. This led to the convenient behavior that all eigenvalues are centered around a single point on the negative real axis, following Robert May's circular law. Without that assumption, the eigenvalue distributions would still be restricted by circles, but now follow the less restrictive Gerschgorin theorem. Based on the latter, the radius scales as $\sim N$, and not $\sim\sqrt{N}$ as in May's circular law. This leads not only to larger circles but also to unions of overlapping circles, restricting the real parts less efficiently \cite{Pazuki_2024}. Using the same degradation rates also does not allow for flexibility in the levels of self-activation and inhibition often observed in biological pathways. Choosing degradation rates from a distribution introduces correlations among corresponding off-diagonal matrix elements $J_{ij}$ and $J_{ji}$ and hence different shapes of eigenvalue distributions \cite{allesina_stabilitycomplexity_2015}. 

 {Due to the optimal Turing network size for relatively small $N$, we did not need to investigate very large matrices. However, to speed up the investigation of large random Jacobians, the concept of reactivity can be included \cite{NEUBERT20021}. Reactivity, describing the transient growth potential of local perturbations in absence of diffusion, could provide a preliminary filter that avoids the full diagonalization of the system's Jacobian. This is because the reactivity is determined by the largest eigenvalue of the Hermitian part of the Jacobian matrix, which is computationally cheaper to compute than the eigenvalues of the full Jacobian matrix.} 

Our work opens new avenues for exciting research in systems biology. In future work, more specific small networks could be investigated similar to the ones shown in Fig. \ref{fig:fig1}, to better understand the distributions of Jacobian matrix elements with pattern-forming capability from sampled parameter values. In addition, examining random networks would allow one to obtain joint distributions of matrix elements and hence correlations among these, leading to renewed investigations into the role of network topology \cite{diego_key_2018}. Of course, more diffusing species could be included, and the effects of boundary conditions investigated. The latter effect was removed in our study by considering systems of infinite domain size, in line with previous studies \cite{scholes_comprehensive_2019}.

 {Another interesting future direction is to study random Turing networks on graphs, moving from continuous to discrete topologies, with diffusion replaced by hopping along the graph's edges \cite{othmer_instability_1971} \cite{nakao_turing_2010} \cite{kouvaris_pattern_2015} \cite{muolo_turing_2024}. Due to more versatile interactions, robustness may be increased. While classical Turing patterns are associated with regular spatial structures, on graphs, the lack of a physical notion of space makes patterns inherently dependent on the network's topology. Patterns may appear irregular on random graphs or when nodes are visualized arbitrarily, but they reflect the underlying eigenmodes of the graph's Laplacian and the topology's influence on reaction-diffusion dynamics. The challenge lies in interpreting these patterns in the context of the system being modeled, as their biological or functional meaning may not rely on spatial regularity but rather on network-specific features. However, applications are vast, ranging from biochemical reactors and neuronal networks to tissues of living cells, synthetic systems, and patchy predator–prey ecosystems.}

In conclusion, we find that Turing patterns do not only occur more frequently by chance than previously thought, but also that there is a surprising sweet spot of network size for most robust Turing patterns. This optimal network size occurred via a stability and instability tradeoff, which did not change significantly when varying the sparsity of the network. This finding supports the idea that the Turing mechanism, if employed by evolution in developmental pathways, is represented by relatively small and hence likely identifiable modules. Understanding their embedding and hierarchical organization in large developmental networks will be worthwhile endeavors.

\newpage
\section*{Methods}

\subsection*{Parameter searches including Latin hypercube sampling}  {Parameter sampling was done with Latin hypercube sampling, which is a statistical method for generating a sample of parameter values from a multidimensional space, ensuring that each parameter is evenly sampled across its range by dividing the space into equal probability intervals \cite{mckay_comparison_2000} \cite{tica_three-node_2023}. Its strength lies in efficiently covering the input space with fewer samples compared to simple random sampling, reducing variance while maintaining representative coverage.} Specifically, for Fig. \ref{fig:fig3} we sampled $10^7$ parameter combinations for each network topology using the {\it lhs} function from {\it pyDOE} library, using a loguniform distribution with the following bounds:  $V_i:\ (0.1, 10)$, $K_{ij}:\ (0.01, 1)$, $b_i:\ (0.001, 0.1)$, $\mu_i:\ (0.01, 1)$ (see Eq. \ref{eq:f_i}). The diffusion constants for Species 1 and 2 were fixed to $1$ and $10$, and the Hill coefficient was set to $n_{ij}=2$ throughout. For each parameter set we calculated the unique steady states using $10$ different initial guess using the Newton-Raphson method (function {\it root} with 'hybrid' setting). In Figs. \ref{fig:fig6} and \ref{fig:fig7} we used $10^3$ random matrices for each $N$ and $\gamma$ pair, and in Fig. \ref{fig:fig8} we used $10^3$ random matrices for each $D_1$ and $D_2$ pair.

\subsection*{Linear stability analysis} 
For each steady state of a parameter set, and there can be more than one due to multistability, we conducted linear stability analysis by calculating and diagonalizing the Jacobian matrices for wave numbers from $0$ to $k_{max}=10$ using stead size $\Delta k=0.01$ and with {\it linalg.eigenvals} function from {\it numpy} library. Subsequently, we analyzed for Turing I, II etc using a simplified version of the classification scheme in \cite{scholes_comprehensive_2019}.

\subsection*{Numerical implementation and code availability} For Fig. \ref{fig:fig3}, the high-dimensional parameter space created by nested loops was flattened, and ran in parallel by dividing up the linear parameter array equally on different cores using the {\it multiprocessing} library. The code used for producing Figs. \ref{fig:fig2}-\ref{fig:fig8} can be accessed on\\ $github.com/Endres\-group/Random\_Turing\_networks$.

\subsection*{Data availability} The datasets generated and analyzed in Figs. \ref{fig:fig3} and \ref{fig:fig4} are available in the Zenodo repository\\ {\it https://zenodo.org/records/13142658 }

\subsection*{Fitting of beta distribution} Fitting a beta distribution to the data of the Jacobian matrix elements in Fig. \ref{fig:fig3} involves estimating the distribution's parameters, typically using the method of moments or maximum likelihood estimation (MLE). The method of moments involves calculating the sample mean and variance to derive initial estimates for the shape parameters $\alpha$ and $\beta$. MLE, on the other hand, maximizes the log-likelihood function, which measures how well the beta distribution with specific parameters explains the observed data, to find the best-fitting parameters. Both methods aim to provide a beta distribution that accurately represents the underlying data distribution. Here, we used the method of moments to obtain an initial guess for the parameter values of $\alpha$ and $\beta$, which we then used as an initial guess for the maximization of the log-likelihood function.

\subsection*{Mathematical proofs} 

\noindent Proof that the outliers of the Jacobian matrices have extreme negative real parts, following approximately $-N-k^2\sum_{i=1}^ND_i$.
This is because the trace is given by $Tr(\bm{J})=Tr(\bm{J_0})-k^2Tr(\bm{D})=-Tr(\bm{1})-k^2Tr(\bm{D}) = -N-k^2\sum_{i=1}^ND_i = \sum_{i=1}^N \lambda_i$, which is invariant under similarity transformations during diagonalization and hence equals the sum of all the eigenvalues $\lambda_i$ ($i=1,\dots, N$). Or in other words, the mean of the eigenvalues stays the same under such a transformation. Hence, at least one eigenvalue, if real, or a complex conjugate pair, if with an imaginary part, becomes an outlier and tends to minus infinity for increasing wave number $k$ at a given network size $N$. This is relevant for Fig. \ref{fig:fig5}A.

Proof that there are no Turing instabilities for random matrices with $N=2$: Based on Eq. \ref{eq:jacobian_with_diff}, the $k$-dependent Jacobian matrix is given by
\begin{equation}
       J= \begin{pmatrix}
        -1-D_1k^2 & g_{12} \\
        g_{21} & -1-D_1k^2 \\ \label{eq:J2D}
        \end{pmatrix}
\end{equation}
Hence, the trace is always negative for all $k$, i.e.
\begin{equation}
Tr_k=-2-(D_1+D_2)k^2<0,
\end{equation}
fulfilling one of Turing's conditions. Furthermore, the determinant without diffusion ($k=0$) is required to be positive, and hence must be
\begin{equation}
Det_0=1-g_{12}g_{21}>0,
\end{equation}
leading to the required stability in absence of diffusion. However, with diffusion ($k>0$), the determinant cannot become negative as needed for a saddle node and hence instability. Instead, we have
\begin{eqnarray}
Det_k&=&(1+D_1k^2)(1+k^2D_2)-g_{12}g_{21}\\
&=&Det_0+(D_1+D_2)k^2+D_1D_2k^4>0.
\end{eqnarray}
Hence, the last of Turing's conditions cannot be fulfilled for two nodes, which is relevant for Figs. \ref{fig:fig7}A and \ref{fig:fig8}A.  {Note, Jacobian matrix Eq. \ref{eq:J2D} is different from Turing's original linear model, in which the diagonal matrix element of the activator is positive without diffusion \cite{turing_chemical_1952}. In our case, all diagonal matrix elements are $-1$, leading to increased stability.}\\

\subsection*{Acknowledgements} We thank Martina Oliver Huidobro, Mark Isalan, David Lacoste, and Roozbeh Pazuki for helpful discussions, and funding through a studentship from the Department of Life Sciences at Imperial College London (to AMG) and the AI-4-EB Consortium for Bioengineered Cells and Systems (BBSRC award BB/W013770/1) (to RGE).

\newpage

\bibliography{references}

\end{document}